\documentstyle[aps,prl,twocolumn,epsfig]{revtex}

\catcode`\@=11
\def\abstract#1{\gdef\@abstract{
\begin{center}
\parbox{14cm}
{\small\rm\ignorespaces#1\par}
\end{center}}}
\def\@maketitle{%
\@preprint
\@title
\ifdim\prevdepth=-1000pt \prevdepth0pt\fi
\@authoraddress
\@date
\@abstract}
\def\section{\@startsection{section}{1}{\z@}{3.5ex plus 1ex minus .2ex}
{2.3ex plus .2ex}{\large\bf}}
\def\thesection{\arabic{section}.}

\def\appendix{\setcounter{section}{0}
\def\thesection{Appendix \Alph{section}:}
\def\theequation{\Alph{section}.\arabic{equation}}}
\def\@citex[#1]#2{\if@filesw\immediate\write\@auxout{\string\citation{#2}}\fi
  \def\@citea{}\@cite{\@for\@citeb:=#2\do
    {\@citea\def\@citea{,\penalty\@m}\@ifundefined
       {b@\@citeb}{{\bf ?}\@warning
       {Citation `\@citeb' on page \thepage \space undefined}}%
\hbox{\csname b@\@citeb\endcsname}}}{#1}}
\def\citer{\@ifnextchar
[{\@tempswatrue\@citexr}{\@tempswafalse\@citexr[]}}
\def\@citexr[#1]#2{\if@filesw\immediate\write\@auxout{\string\citation{#2}}\fi
  \def\@citea{}\@cite{\@for\@citeb:=#2\do
    {\@citea\def\@citea{--\penalty\@m}\@ifundefined
       {b@\@citeb}{{\bf ?}\@warning
       {Citation `\@citeb' on page \thepage \space undefined}}%
\hbox{\csname b@\@citeb\endcsname}}}{#1}}
\relax

\begin{document}

\thispagestyle{empty}

\begin{titlepage}

\begin{flushright}
CERN-TH/99-164\\
hep-ph/9906274
\end{flushright}

\vspace{3cm}
\begin{center}
\boldmath
\Large\bf Searching for New Physics in Non-Leptonic $B$ Decays
\unboldmath
\end{center}

\vspace{1.2cm}
\begin{center}
\large Robert Fleischer\, and\, Joaquim Matias\\[0.1cm]
\large{\sl Theory Division, CERN, CH-1211 Geneva 23, Switzerland}
\end{center}

\vspace*{1.7cm}

\begin{center}
\large{\bf Abstract}

\vspace*{0.6cm}

\begin{tabular}{p{13truecm}}
We present  allowed regions in the space of observables of certain 
non-leptonic $B$-meson decays that characterize these modes within 
the Standard Model. 
A future measurement of observables lying significantly outside of
these regions would  indicate the 
presence of new physics. Making use of $SU(3)$ arguments, we give 
the range for $B\to\pi K$ decays, 
and  for the system of $B_d\to\pi^+\pi^-$,
$B_s\to K^+K^-$  modes.
\end{tabular}

\end{center}

\vspace*{4truecm}

\noindent
CERN-TH/99-164\\
June 1999

\end{titlepage}

\newpage
\thispagestyle{empty}
\parbox{17cm}
{\mbox{}}

\newpage
\thispagestyle{empty}
\parbox{17cm}
{\mbox{}}

\newpage
\setcounter{page}{1}

\title
{Searching for New Physics in Non-Leptonic $B$ Decays}

\author{Robert Fleischer, Joaquim Matias}

\address{Theory Division, CERN, CH-1211 Geneva 23, Switzerland}

\abstract{We present  allowed regions in the space of observables of certain 
non-leptonic $B$-meson decays that characterize these modes within 
the Standard Model. 
A future measurement of observables lying significantly outside of
these regions would  indicate the 
presence of new physics. Making use of $SU(3)$ arguments, we give 
the range for $B\to\pi K$ decays, 
and  for the system of $B_d\to\pi^+\pi^-$,
$B_s\to K^+K^-$  modes. \\
\vskip-5pt PACS numbers: 13.25.Hw, 11.30.Er, 11.30.Hv\\[-1.2cm]}
\maketitle

%\narrowtext

As is well known, the $B$-meson system provides a very fertile testing 
ground for the Standard-Model description of CP violation, where this 
phenomenon originates from a complex phase in the 
Cabibbo--Kobayashi--Maskawa matrix (CKM matrix). In order to search for 
new physics, one of the main methods is to overconstrain the three angles 
$\alpha$, $\beta$ and $\gamma$ of the usual non-squashed unitarity triangle 
of the CKM matrix, thereby searching for possible discrepancies. During 
the last years, many interesting strategies were proposed to accomplish 
this task \cite{revs}.

In this paper, we propose a simple approach, which offers the exciting 
possibility of immediate indications of new physics at future $B$-decay 
experiments. It relies on the fact that certain non-leptonic $B$-meson 
decays into two light pseudoscalar mesons can be characterized, within
the Standard Model (SM), by regions arising in the space of the 
corresponding observables. If future measurements of these observables 
should result in values lying significantly outside of these regions, we 
would have an indication for the presence of new physics. 

We show these regions for two different combinations of $B\to\pi K$
modes \cite{grl}--\cite{Neubert}, as well as for the system of 
$B_d\to\pi^+\pi^-$ and $B_s\to K^+K^-$ decays \cite{BsKK}. In order to 
evaluate them, we have to make use of $SU(3)$ flavour-symmetry arguments in 
both cases. In the $B\to\pi K$ case, which is very promising for 
$e^+$--$\,e^-$ $B$-factories, an additional dynamical assumption concerning 
final-state-interaction (FSI) effects has to be made \cite{FSI}. This is 
not necessary in the $B_d\to\pi^+\pi^-$, $B_s\to K^+K^-$ system, 
which is ideally suited for ``second-generation'' $B$-physics experiments 
at hadron machines, such as LHCb or BTeV. Since flavour-changing 
neutral-current ``penguin'' processes play an important role in $B\to\pi K$, 
$B_d\to\pi^+\pi^-$ and $B_s\to K^+K^-$ decays, they may well be affected 
by new physics \cite{Neubert,new-phys,GNK}. Moreover, the unitarity of the 
CKM matrix is used to evaluate the corresponding allowed regions. 

Let us turn to the $B\to\pi K$ system first, which already allows us
to confront the contours in the space of observables with experimental
data from the CLEO collaboration \cite{cleo}. We will consider two 
different combinations of $B\to\pi K$ decays: the charged modes 
$B^\pm\to\pi^\pm K$ and $B^\pm\to\pi^0K^\pm$ \cite{grl,NR,Neubert}, and the 
``mixed'' combination $B^\pm\to\pi^\pm K$, $B_d\to\pi^\mp K^\pm$ 
\cite{BpiK-mix}. Within the SM, we have
\begin{equation}
P\equiv A(B^+\to\pi^+K^0)\propto \left[1+\rho\,e^{i\vartheta}e^{i\gamma}
\right]{\cal P}_{tc}\,,
\end{equation}
where
\begin{equation}
\rho\,e^{i\vartheta}=\frac{\lambda^2R_b}{1-\lambda^2/2}
\left[1-\left(\frac{{\cal P}_{uc}+{\cal A}}{{\cal P}_{tc}}\right)\right],
\end{equation}
with $\lambda\equiv |V_{us}|=0.22$, $A\equiv|V_{cb}|/\lambda^2=0.81\pm0.06$ 
and $R_b\equiv|V_{ub}/(\lambda V_{cb})|=0.41\pm0.07$. The amplitudes 
${\cal A}$ and ${\cal P}_{tc}\equiv|{\cal P}_{tc}|e^{i\delta_{tc}}$ 
(${\cal P}_{uc}$) are due to annihilation and penguin topologies with 
internal top- and charm-quark (up- and charm-quark) exchanges, respectively. 
The $SU(2)$ isospin symmetry of strong interactions implies 
\begin{eqnarray}
A(B^+\to\pi^+K^0)&+&\sqrt{2}\,A(B^+\to\pi^0K^+)\nonumber\\
&=&-\left[(T+C)\,+\,P_{\rm ew}\right],
\end{eqnarray}
where the amplitudes
\begin{equation}
T+C\equiv|T+C|\,e^{i\delta_{T+C}}\,e^{i\gamma}\,\,\mbox{and}\,\,\,
P_{\rm ew}=-\,|P_{\rm ew}|e^{i\delta_{\rm ew}}
\end{equation}
arise from current--current and electroweak penguin operators, 
respectively (the $\delta$s denote strong phases). The $SU(3)$ flavour 
symmetry of strong interactions allows us to fix $|T+C|$ with the help 
of the decay $B^+\to\pi^+\pi^0$ \cite{grl}:
\begin{equation}\label{rc-det}
T+C=-\,\sqrt{2}\,\frac{V_{us}}{V_{ud}}\,
\frac{f_K}{f_{\pi}}\,A(B^+\to\pi^+\pi^0)\,,
\end{equation}
where the kaon and pion decay constants take into account factorizable
$SU(3)$-breaking corrections. Moreover, we have in the strict $SU(3)$ limit
\cite{NR}
\begin{equation}\label{EWP}
\left|\frac{P_{\rm ew}}{T+C}\right|\,
e^{i(\delta_{\rm ew}-\delta_{T+C})}=0.66\times
\left[\frac{0.41}{R_b}\right].
\end{equation}
The factorizable $SU(3)$-breaking corrections to this relation are very 
small, and its theoretical accuracy is only limited by non-factorizable
effects. In a recent paper \cite{fact}, an interesting approach making 
use of a heavy-quark expansion for non-leptonic $B$ decays was proposed 
that could help to reduce these uncertainties.

The decays $B^+\to\pi^+K^0$ and $B^+\to\pi^0K^+$ provide the following
observables:
\begin{eqnarray}
R_{\rm c}&\equiv&2\left[\frac{\mbox{BR}(B^+\to\pi^0K^+)+
\mbox{BR}(B^-\to\pi^0K^-)}{\mbox{BR}(B^+\to\pi^+K^0)+
\mbox{BR}(B^-\to\pi^-\overline{K^0})}\right]\\
A_0^{\rm c}&\equiv&2\left[\frac{\mbox{BR}(B^+\to\pi^0K^+)-
\mbox{BR}(B^-\to\pi^0K^-)}{\mbox{BR}(B^+\to\pi^+K^0)+
\mbox{BR}(B^-\to\pi^-\overline{K^0})}\right],
\end{eqnarray}
where the factor of 2 has been introduced to absorb the normalization factor
of the $\pi^0$. The present CLEO data imply $R_{\rm c}=1.3\pm0.5$ 
\cite{cleo}; very recently, also the first results for CP-violating 
asymmetries in charmless hadronic $B$-meson decays were reported, leading 
to $A_0^{\rm c}=0.35\pm0.34$. 

In order to parametrize $R_{\rm c}$ and $A_0^{\rm c}$, it is useful to 
introduce 
\begin{equation}
r_{\rm c}\equiv\frac{|T+C|}{\sqrt{\langle|P|^2\rangle}},\quad q\,e^{i\omega}
\equiv\left|\frac{P_{\rm ew}}{T+C}\right|\,
e^{i(\delta_{\rm ew}-\delta_{T+C})}\,.
\end{equation}
The general expressions for $R_{\rm c}$ and $A_0^{\rm c}$ in terms of
these parameters and $\rho\,e^{i\vartheta}$ can be found in \cite{BF}. 
Here we restrict ourselves, for simplicity, to the case of $\rho=0$, 
corresponding to the neglect of rescattering processes \cite{FSI}, 
and to $\omega=0$, corresponding to (\ref{EWP}). Then we obtain
\begin{eqnarray}
R_{\rm c}&=&1-2\,r_{\rm c}\left(\cos\gamma-q\right)\cos\delta_{\rm c}+
v^2r_{\rm c}^2\label{Rc-expr}\\
A_0^{\rm c}&=&2\,r_{\rm c}\sin\delta_{\rm c}\sin\gamma,\label{A0-expr}
\end{eqnarray}
where $\delta_{\rm c}\equiv\delta_{T+C}-\delta_{tc}$ and $v\equiv
\sqrt{1-2\,q\cos\gamma+q^2}$. Since $r_{\rm c}$ and $q$ can be fixed 
through (\ref{rc-det}) and (\ref{EWP}), respectively, the two 
observables $R_{\rm c}$ and $A_0^{\rm c}$ depend on the two ``unknowns'' 
$\delta_{\rm c}$ and $\gamma$. Consequently, if we fix $r_{\rm c}$ and 
$q$ -- present data give $r_{\rm c}=0.21\pm0.06$ and $q=0.63\pm0.15$ -- 
and vary $\delta_{\rm c}$ and $\gamma$ within $[0^\circ, 
360^\circ]$, (\ref{Rc-expr}) and (\ref{A0-expr}) imply an allowed region 
in the $R_{\rm c}$--$A_0^{\rm c}$ plane.

%%%%%%%%%%%%%%%%%%%%%%%%%%%%%%%%%%%%%%%%%%%%%%%%%%%%%%%%%%%%%%%%%%%
\begin{figure}
\vspace{-2.8cm}
\begin{tabular}{c}
\vspace{-5cm}
   \epsfysize=9.5cm
   \epsffile{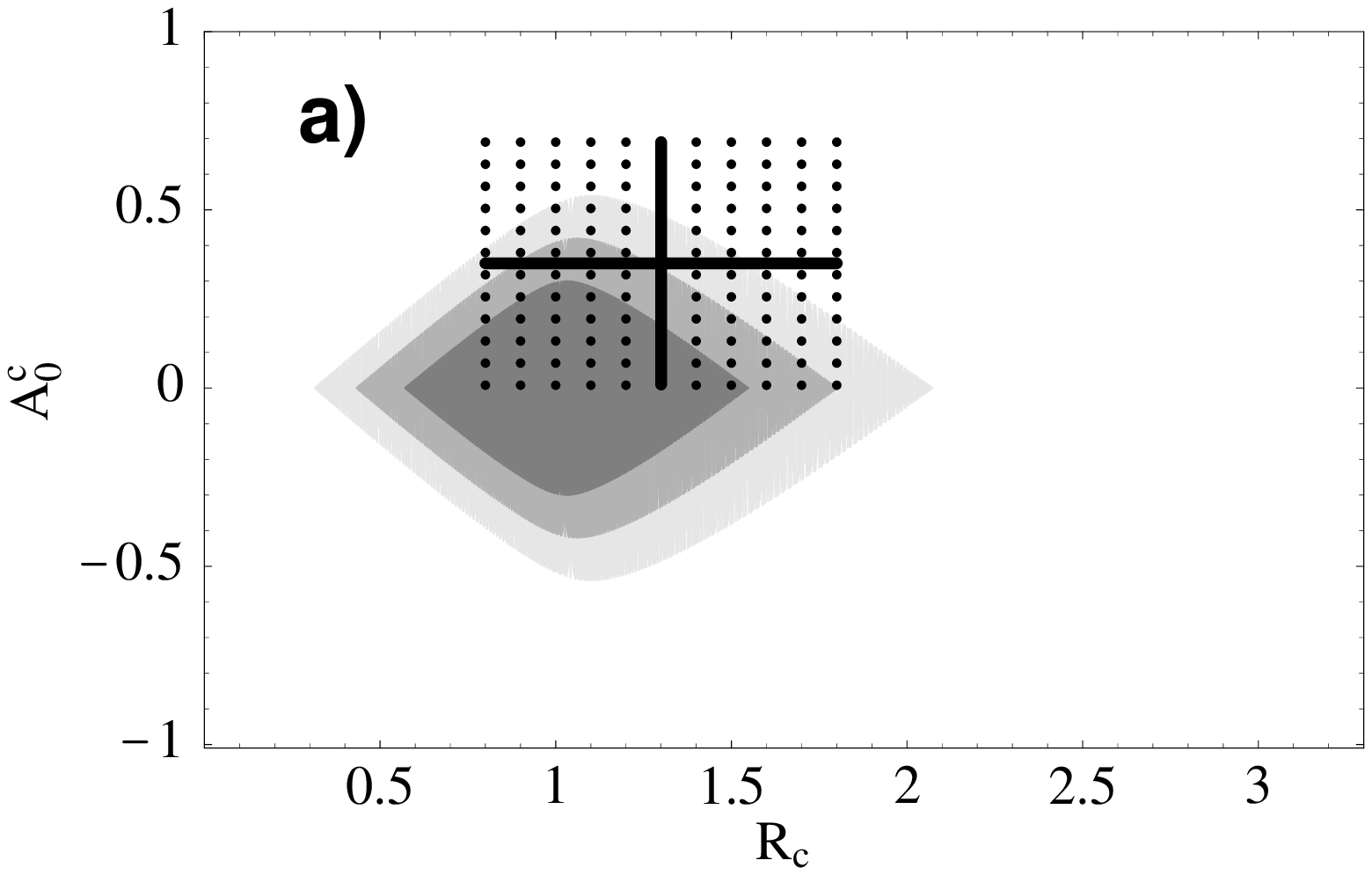}
\\
\vspace{-4.6cm}
   \epsfysize=9.5cm
   \epsffile{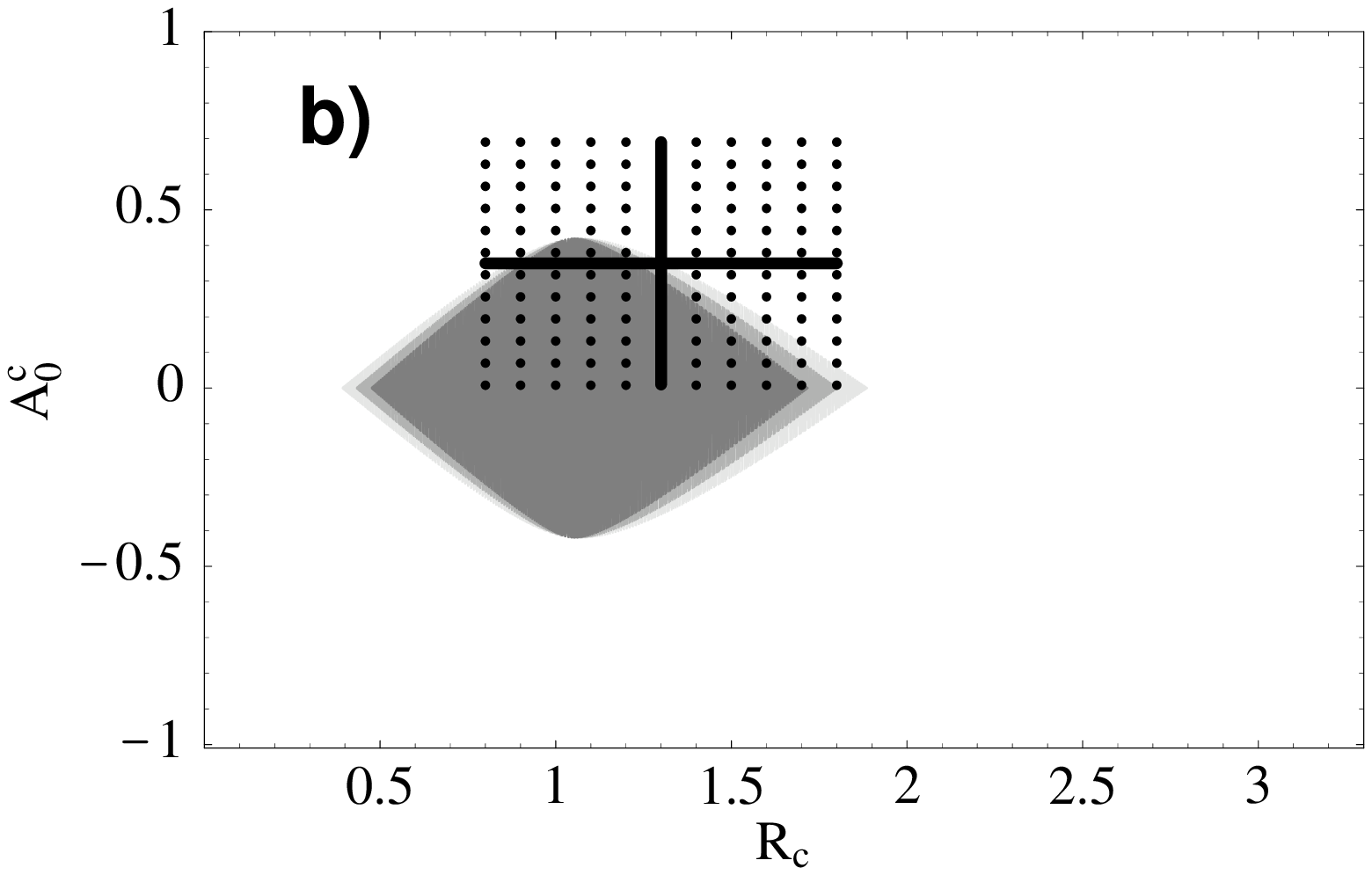}
\vspace{2.2cm}
\end{tabular}
\caption[dummy]{\small Allowed region in the $R_{\rm c}$--$A_0^{\rm c}$ 
plane, characterizing $B^\pm\to\pi^\pm K$, $\pi^0K^\pm$ in the
SM: (a) $0.15\leq r_{\rm c}\leq0.27$, $q=0.63$; (b) $r_{\rm c}=0.21$, 
$0.48\leq q\leq0.78$. FSI effects are neglected.
\label{fig1}}
\end{figure}
%%%%%%%%%%%%%%%%%%%%%%%%%%%%%%%%%%%%%%%%%%%%%%%%%%%%%%%%%%%%%%%%%%%

\begin{figure}
\vspace{-2.7cm}
   \epsfysize=10cm
   \centerline{\epsffile{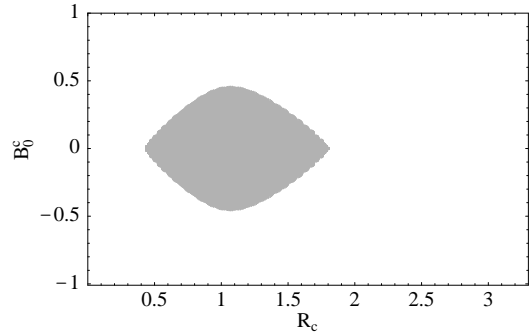}}
\vspace{-2.4cm}
\caption[dummy]{\small Allowed region in the $R_{\rm c}$--$B_0^{\rm c}$ 
plane, characterizing $B^\pm\to\pi^\pm K$, $\pi^0K^\pm$ in the
SM in the presence of large FSI effects, which are described by 
$\rho=0.15$ ($r_{\rm c}=0.21$, $q=0.63$).
\label{fig-FSI}}
\end{figure}

In Fig.\ \ref{fig1}, we show this region for the currently allowed values 
of the parameters $r_{\rm c}$ and $q$. The small dependence on the latter 
parameter (see Fig.\ \ref{fig1}(b)) is due to the suppression through 
$r_{\rm c}$ in (\ref{Rc-expr}). A similar suppression is also effective
for the terms of ${\cal O}(\rho)$  in $R_{\rm c}$, which are related to FSI
effects. If we use the observable
\begin{eqnarray}
B_0^{\rm c}\equiv&A_0^{\rm c}-\left[\frac{\mbox{BR}(B^+\to\pi^
+K^0)-
\mbox{BR}(B^-\to\pi^-\overline{K^0})}{\mbox{BR}(B^+\to\pi^+K^0)+
\mbox{BR}(B^-\to\pi^-\overline{K^0})}\right]
\end{eqnarray}
instead of $A_0^{\rm c}$, the terms of ${\cal O}(\rho)$ are 
suppressed by $r_{\rm c}$ as well, as was also noted in Ref.\ \cite{Neubert}. 
In the case of Fig.~\ref{fig1}, the FSI effects are neglected, leading to 
$B_0^{\rm c}=A_0^{\rm c}$. If we choose $r_c=0.21$, $q=0.63$, and 
assume that $\rho=0.15$, which would correspond to very large rescattering 
effects, while keeping $\vartheta\in[0^\circ, 360^\circ]$ as a free 
parameter, we obtain the allowed region shown in Fig.\ \ref{fig-FSI}. 
This figure shows nicely that the impact of FSI effects on the allowed 
region in the $R_{\rm c}$--$B_0^{\rm c}$ plane is very small. Let 
us nevertheless note that the FSI effects can be probed -- 
and in principle even included in Fig.\ \ref{fig1} -- with the help of 
additional experimental data \cite{BF,FSI-cont}, for example on 
$B^\pm\to K^\pm K$ modes. 

%%%%%%%%%%%%%%%%%%%%%%%%%%%%%%%%%%%%%%%%%%%%%%%%%%%%%%%%%%%%%%%%%%%%%
%%%%%%%%%%%%%%%%%%%%%%%%%%%%%%%%%%%%%%%%%%%%%%%%%%%%%%%%%%%%%%%%%%%

The dotted range in Fig.\ \ref{fig1} corresponds to the present CLEO results 
for $R_{\rm c}$ and $A_0^{\rm c}$. If future measurements of $R_{\rm c}$ and 
$A_0^{\rm c}$ should give values lying significantly outside the allowed 
region shown in Fig.\ \ref{fig1}, we would have an indication for new 
physics. On the other hand, if we should find values lying inside 
this region, this would not automatically imply a ``confirmation'' of the 
SM. In this case, it would be possible to extract a value of 
$\gamma$ by following the strategies proposed in \cite{NR,BF}, which may 
well lead to discrepancies with the values of $\gamma$ that are implied 
by theoretically clean strategies, using pure ``tree'' decays, such as 
$B\to DK$ or $B_s\to D_s^\mp K^\pm$, or by the usual ``indirect'' fits 
of the unitarity triangle. In a recent paper \cite{GNK}, several specific 
models were employed to expore the impact of new physics on $B\to\pi K$
decays. For example, in models with an extra $Z'$ boson or in SUSY models
with broken $R$-parity, the resulting electroweak penguin coefficients can
be much larger than in the SM, since they arise already at the tree level. 
In this paper, it is not our purpose to consider specific models for new
physics. However, we plan to come back to this issue in a forthcoming 
publication.

%%%%%%%%%%%%%%%%%%%%%%%%%%%%%%%%%%%%%%%%%%%%%%%%%%%%%%%%%%%%%%%%%%%%%
\begin{figure} 
\vspace{-2.7cm}
   \epsfysize=10cm
   \centerline{\epsffile{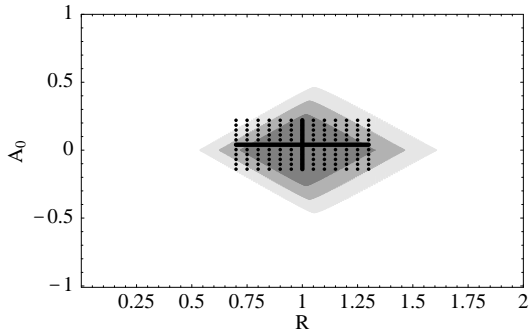}}
\vspace{-2cm}
\caption[dummy]{\small Allowed region in the $R$--$A_0$ plane, characterizing 
$B^\pm\to\pi^\pm K$, $B_d\to\pi^\mp K^\pm$ within the SM for 
$0.13\leq r\leq0.23$, $q_{\rm C}\,e^{i\omega_{\rm C}}=0.66\times0.25$. 
FSI effects are neglected.
\label{fig2}}
\end{figure}  
%%%%%%%%%%%%%%%%%%%%%%%%%%%%%%%%%%%%%%%%%%%%%%%%%%%%%%%%%%%%%%%%%%%

In Fig.\ \ref{fig2}, we show the allowed region for the observables of 
the $B^\pm\to\pi^\pm K$, $B_d\to\pi^\mp K^\pm$ system \cite{BpiK-mix}, 
where $R$ and $A_0$ correspond to $R_{\rm c}$ and $A_0^{\rm c}$, 
respectively; explicit expressions can be found in \cite{BF}, where also 
the parameters $r$ and $q_{\rm C}\,e^{i\omega_{\rm C}}$ are defined 
properly. The latter describes ``colour-suppressed'' electroweak penguins,
which are usually expected to play a minor role \cite{EWPs}. In contrast to 
the charged case, $r$ and $q_{\rm C}\,e^{i\omega_{\rm C}}$ cannot be fixed 
by using only flavour-symmetry arguments. To this end, we have to employ, 
in addition, certain dynamical assumptions, such as arguments involving 
the ``factorization'' hypothesis, and have to keep in mind that the
parameters thus determined  may also be affected by FSI effects, which have 
been neglected in Fig.\ \ref{fig2}. However, there are important 
experimental indicators for such rescattering processes, for example the
branching ratios of $B\to K K$ modes or a sizeable direct CP asymmetry in 
$B^\pm\to\pi^\pm K$. In order to reduce these uncertainties, also the 
approach proposed in Ref.\ \cite{fact} may turn out to be very useful. 
The dotted range in Fig.\ \ref{fig2} represents the present CLEO results 
$R=1.0\pm0.3$ and $A_0=0.04\pm0.18$, which coincides perfectly with the
allowed region implied by the SM. This feature should be compared with the
situation in Fig.\ 1. Unfortunately, the present experimental uncertainties 
are too large to speculate on new-physics effects. However, the experimental 
situation should improve considerably in the next couple of years. 

Let us now focus on the decays $B_d\to\pi^+\pi^-$ and $B_s\to K^+K^-$. The 
latter mode is not accessible at the $e^+$--$\,e^-$ $B$-factories operating 
at the $\Upsilon(4S)$ resonance, but is very promising for 
``second-generation'' $B$-decay experiments at hadron machines. From a
theoretical point of view, the $B_d\to\pi^+\pi^-$, $B_s\to K^+K^-$ system 
has some advantages in comparison with the $B\to\pi K$ approach, as we 
will see below. Within the SM, the $B_d^0\to\pi^+\pi^-$ decay
amplitude can be parametrized, in a completely general way, as 
follows \cite{BsKK}:
\begin{equation}\label{Bd-ampl2}
A(B_d^0\to\pi^+\pi^-)\propto e^{i\gamma}\left[1-d\,e^{i\theta}e^{-i\gamma}
\right],
\end{equation}
where the parameter
\begin{equation}\label{ap-def}
d\,e^{i\theta}\equiv\frac{1}{(1-\lambda^2/2)R_b}
\left(\frac{A_{\rm pen}^{ct}}{A_{\rm cc}^{u}+A_{\rm pen}^{ut}}\right)
\end{equation}
describes -- sloppily speaking -- the ratio of ``penguin'' to ``tree''
contributions. Employing a notation similar to that in (\ref{Bd-ampl2}) 
yields
\begin{equation}\label{Bs-ampl}
A(B_s^0\to K^+K^-)\propto e^{i\gamma}\left[1+\frac{1}{\epsilon}\,
d'e^{i\theta'}e^{-i\gamma}\right],
\end{equation}
where $d'e^{i\theta'}$ corresponds to (\ref{ap-def}), and 
$\epsilon\equiv\lambda^2/(1-\lambda^2)$. The time evolution of the decay
$B_s\to K^+K^-$ provides the following time-dependent CP asymmetry: 
\begin{eqnarray}
\lefteqn{a_{\rm CP}(t)\equiv\frac{\Gamma(B^0_s(t)\to f)-
\Gamma(\overline{B^0_s}(t)\to f)}{\Gamma(B^0_s(t)\to f)+
\Gamma(\overline{B^0_s}(t)\to f)}}\nonumber\\
&&~~~=\frac{2\,e^{-\Gamma_s t}\left[{\cal A}_{\rm CP}^{\rm dir}
\cos(\Delta M_s t)+{\cal A}_{\rm CP}^{\rm mix}\sin(\Delta M_s t)\right]}{
e^{-\Gamma_{\rm H}^{(s)}t}+e^{-\Gamma_{\rm L}^{(s)}t}+
{\cal A}_{\rm \Delta\Gamma}\left(e^{-\Gamma_{\rm H}^{(s)}t}-
e^{-\Gamma_{\rm L}^{(s)}t}\right)},
\end{eqnarray}
where ${\cal A}_{\rm CP}^{\rm dir}$, ${\cal A}_{\rm CP}^{\rm mix}$ and
${\cal A}_{\rm \Delta\Gamma}$ satisfy the relation
\begin{equation}\label{obs-rel}
\bigl({\cal A}_{\rm CP}^{\rm dir}\bigr)^2+\bigl({\cal A}_{\rm CP}^{\rm mix}
\bigr)^2+\bigl({\cal A}_{\rm \Delta\Gamma}\bigr)^2=1.
\end{equation}
Using (\ref{Bs-ampl}), we obtain \cite{BsKK}
\begin{equation}\label{ACP-dir-s}
{\cal A}_{\rm CP}^{\rm dir}(B_s\to K^+K^-)=
\frac{2\,\tilde d'\sin\theta'\sin\gamma}{1+
2\,\tilde d'\cos\theta'\cos\gamma+\tilde d'^2}
\end{equation}
\begin{eqnarray}
\lefteqn{{\cal A}_{\rm CP}^{\rm mix}(B_s\to K^+K^-)=}\label{ACP-mix-s}\\
&&~~~~\frac{\sin(\phi_s+2\gamma)+2\,\tilde d'
\cos\theta'\sin(\phi_s+\gamma)+
\tilde d'^2\sin\phi_s}{1+2\,\tilde d'\cos\theta'\cos\gamma+
\tilde d'^2},\nonumber
\end{eqnarray}
where $\tilde d'\equiv d'/\epsilon$, and $\phi_s\equiv-2\delta\gamma=
2\,\mbox{arg}(V_{ts}^\ast V_{tb})$ denotes the $B^0_s$--$\overline{B^0_s}$ 
mixing phase. Within the SM, we have $2\delta\gamma\approx0.03$ 
due to a Cabibbo suppression of ${\cal O}(\lambda^2)$, implying that 
$\phi_s$ is very small. 

The expression for the time-dependent $B_d\to\pi^+\pi^-$ CP asymmetry 
simplifies considerably, since the width difference $\Delta\Gamma_d\equiv
\Gamma_{\rm H}^{(d)}-\Gamma_{\rm L}^{(d)}$ between the $B_d$ mass eigenstates
is -- in contrast to the expected situation in the $B_s$ system -- 
negligibly small. Using (\ref{Bd-ampl2}), the corresponding CP-violating 
observables can be expressed as \cite{BsKK}
\begin{equation}
{\cal A}_{\rm CP}^{\rm dir}(B_d\to\pi^+\pi^-)=
-\left[\frac{2\,d\sin\theta\sin\gamma}{1-
2\,d\cos\theta\cos\gamma+d^2}\right]\label{ACP-dir-d}
\end{equation}
\begin{eqnarray}
\lefteqn{{\cal A}_{\rm CP}^{\rm mix}(B_d\to\pi^+\pi^-)=}\label{ACP-mix-d}\\
&&~~~~\frac{\sin(\phi_d+2\gamma)-2\,d\,\cos\theta\,\sin(\phi_d+\gamma)+
d^2\sin\phi_d}{1-2\,d\cos\theta\cos\gamma+d^2},\nonumber
\end{eqnarray}
where $\phi_d=2\beta$ denotes the $B^0_d$--$\overline{B^0_d}$ mixing phase.
It should be emphasized that (\ref{ACP-dir-s}), (\ref{ACP-mix-s}) and 
(\ref{ACP-dir-d}), (\ref{ACP-mix-d}) are completely general parametrizations 
within the SM, taking also into account all kinds of penguin and
FSI effects.  

Since the decays $B_d\to\pi^+\pi^-$ and $B_s\to K^+K^-$ are related
to each other by interchanging all strange and down quarks, the $U$-spin
flavour symmetry implies
\begin{equation}\label{U-rel}
d'e^{i\theta'}=d\,e^{i\theta}.
\end{equation}
Interestingly, this relation is not affected by $U$-spin-breaking 
corrections within a modernized version of the ``Bander--Silverman--Soni''
mechanism \cite{bss}, which relies -- among other things -- also on the 
``factorization'' hypothesis \cite{BsKK}. Consequently, unless 
non-factorizable effects should have a dramatic impact, the $U$-spin-breaking 
corrections to (\ref{U-rel}) are probably moderate. We are optimistic
that future $B$-decay experiments will also provide valuable insights
into $SU(3)$-breaking effects. Moreover, further work along the 
lines of Ref.\ \cite{fact} may lead to a better theoretical understanding 
of these effects.

%%%%%%%%%%%%%%%%%%%%%%%%%%%%%%%%%%%%%%%%%%%%%%%%%%%%%%%%%%%%%%%%%%%
\begin{figure}[h]
\vspace{-1cm}
   \epsfysize=9.3cm
   \centerline{\epsffile{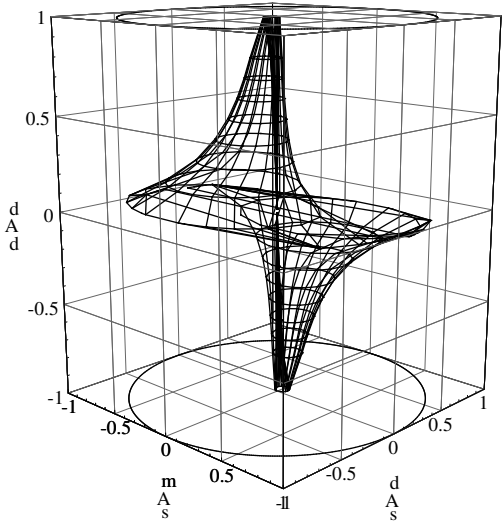}}
\vspace{-1cm}
\caption[dummy]{\small The allowed region in the space of the CP
asymmetries
$A^{\rm d}_s\equiv{\cal A}_{\rm CP}^{\rm dir}(B_s\to K^+K^-)$, 
$A^{\rm m}_s\equiv{\cal A}_{\rm CP}^{\rm mix}(B_s\to K^+K^-)$ and 
$A^{\rm d}_d\equiv{\cal A}_{\rm CP}^{\rm dir}(B_d\to\pi^+\pi^-)$, which 
characterize the $B_s\to K^+K^-$, $B_d\to\pi^+\pi^-$ system within the SM 
($\phi_s=0$).
\label{fig3}}
\end{figure}
%%%%%%%%%%%%%%%%%%%%%%%%%%%%%%%%%%%%%%%%%%%%%%%%%%%%%%%%%%%%%%%%%%%

If we use the $U$-spin relation (\ref{U-rel}), the three observables 
$A^{\rm d}_s\equiv{\cal A}_{\rm CP}^{\rm dir}(B_s\to K^+K^-)$, 
$A^{\rm m}_s\equiv{\cal A}_{\rm CP}^{\rm mix}(B_s\to K^+K^-)$ and 
$A^{\rm d}_d\equiv{\cal A}_{\rm CP}^{\rm dir}(B_d\to\pi^+\pi^-)$ depend 
on the two hadronic parameters $d$ and $\theta$, as well as on the CKM angle 
$\gamma$ and the $B^0_s$--$\overline{B^0_s}$ mixing phase $\phi_s$. 
However, the latter quantity is negligibly small in the SM, 
i.e.\ $\phi_s^{\rm SM}\approx0$. Consequently, if we keep $d$ as a
free parameter, i.e.\ $0\leq d\leq\infty$, and vary $\theta$ and $\gamma$
in the interval $[0^\circ,360^\circ]$, (\ref{ACP-dir-s}),
(\ref{ACP-mix-s}) and 
(\ref{ACP-dir-d}) fix a three-dimensional region in the space of the 
observables $A^{\rm d}_s$, $A^{\rm m}_s$ and $A^{\rm d}_d$, characterizing 
the $B_s\to K^+K^-$, $B_d\to\pi^+\pi^-$ system within the SM. This region 
is shown in Fig.\ \ref{fig3}, where the circles with radius 1 fix a cylinder 
in the $A^{\rm d}_d$ direction, which is due to (\ref{obs-rel}), implying 
$(A^{\rm d}_s)^2+(A^{\rm m}_s)^2\leq1$. An interesting feature of the 
$B_d\to\pi^+\pi^-$, $B_s\to K^+K^-$ predicted region is a hole, which
allows for 
new physics also inside the volume. If one restricts the penguin parameter 
$d$ to be smaller than 1, which seems to be quite plausible, this hole 
would be enlarged. It is also interesting to note that the $U$-spin
flavour 
symmetry implies, within the SM, that the direct CP asymmetries of 
$B_s\to K^+K^-$ and $B_d\to\pi^+\pi^-$ have opposite signs; equal signs 
would be an indication for new physics. In contrast to the $B\to\pi K$ case, 
we do not have to worry about any FSI effects in the $B_d\to\pi^+\pi^-$, 
$B_s\to K^+K^-$ system, and no additional information is required to fix 
certain parameters such as $r_{\rm c}$ or $q$. 

A future measurement of observables lying significantly outside of the 
region shown in Fig.\ \ref{fig3} would be an indication of new physics. 
Such a discrepancy could either be due to CP-violating new-physics 
contributions to $B^0_s$--$\overline{B^0_s}$ mixing, or to the 
$B_d\to\pi^+\pi^-$, $B_s\to K^+K^-$ decay amplitudes. The former case 
would also be indicated simultaneously by large CP-violating effects in 
the mode $B_s\to J/\psi\,\phi$, which would allow us to extract 
the $B^0_s$--$\overline{B^0_s}$ mixing phase $\phi_s$ (see, for example, 
\cite{ddf1}). A discrepancy between the measured $B_d\to\pi^+\pi^-$,
$B_s\to K^+K^-$ observables and the region corresponding to the value 
of $\phi_s$ thus 
determined  would then signal new-physics contributions 
to the $B_d\to\pi^+\pi^-$, $B_s\to K^+K^-$ decay amplitudes. On the other 
hand, if $B_s\to J/\psi\,\phi$ should exhibit negligible CP-violating effects,
any discrepancy between the $B_d\to\pi^+\pi^-$, $B_s\to K^+K^-$
observables
and the volume shown in Fig.~\ref{fig3} would indicate new-physics 
contributions to the corresponding decay amplitudes. On the other hand,
if the observables should lie within the SM predicted region,
we
can extract a value for the CKM angle $\gamma$ by following the strategy
presented in \cite{BsKK}, which may well be in disagreement with those
implied by theoretically clean strategies making use of pure ``tree'' 
decays, thereby also indicating  the presence of new physics. 

If we use the $B^0_d$--$\overline{B^0_d}$ mixing phase $\phi_d$, which 
can be determined, for instance, with the help of the ``gold-plated''
mode $B_d\to J/\psi\,K_{\rm S}$, as an additional input, we may also
fix a three-dimensional region in the space of the observables ${\cal
A}_{\rm CP}^{\rm dir}
(B_d\to\pi^+\pi^-)$, ${\cal A}_{\rm CP}^{\rm mix}(B_d\to\pi^+\pi^-)$
and ${\cal A}_{\rm CP}^{\rm dir}(B_s\to K^+K^-)$ through the Standard-Model
expressions (\ref{ACP-dir-s}), (\ref{ACP-dir-d}) and (\ref{ACP-mix-d}). 
Since the decays $B_s\to K^+K^-$ and $B_d\to\pi^\mp K^\pm$ differ only 
in their spectator quarks, we have
${\cal A}_{\rm CP}^{\rm dir}(B_s\to K^+K^-)\approx{\cal A}_{\rm CP}^{\rm dir}
(B_d\to\pi^\mp K^\pm)$.
Consequently, that figure would also be interesting for the $e^+$--\,$e^-$
$B$-factories, where $B_s\to K^+K^-$ is not accessible. However, we 
should keep it in mind that this relation relies not only on
flavour-symmetry arguments, but also on a certain dynamical input 
concerning ``exchange'' and ``penguin annihilation'' topologies \cite{BsKK}, 
which may be enhanced in the presence of large FSI effects. 

To summarize, we have presented a simple strategy, which may provide
immediate indications for new physics at future $B$-decay experiments.
We plan to discuss in more detail several of the features described
briefly here in a forthcoming paper.

\vspace*{0.3truecm}

\noindent J.M. acknowledges the financial support from a Marie Curie EC 
Grant (TMR-ERBFMBICT 972147).

\end{document}